# Mesoscopic quantized properties of magnetic-dipolar-mode oscillations in disk ferromagnetic particles

E.O. Kamenetskii, R. Shavit, and M. Sigalov

Department of Electrical and Computer Engineering,
Ben Gurion University of the Negev, Beer Sheva, 84105, ISRAEL

**Abstract**

Magnetic dipolar mode or magnetostatic (MS) oscillations in ferrite samples have the wavelength much smaller than the electromagnetic wavelength at the same frequency and, at the same time, much larger than the exchange interaction spin wavelength. This intermediate position between the electromagnetic and spin wave (exchange interaction) processes reveals very special behaviors of the geometrical effects. It was shown recently that magnetic dipolar mode oscillations in a normally magnetized ferromagnetic disk are characterized by discrete energy levels resulting from the structural confinement. In this article we give results of the energy spectra in MS wave ferrite disks taking into account nonhomogeneity of the internal DC magnetic field.

**1. Introduction**

The MS-mode characterization in ferrite samples looks as a relatively straightforward and old problem in magnetism. Nevertheless, some aspects of such oscillations should be re-considered in a view of the macroscopically quantized methods. In last years, there has been a renewed interest in high frequency dynamic properties of finite size magnetic structures. In a series of new publications, confinement phenomena of high-frequency magnetization dynamics in magnetic particles have been the subject of much experimental and theoretical attention (see [1] and references therein). Mainly, these works are devoted to the important studies of the localized spin-wave spectra, but do not focus on the energy eigenstates of a whole ferrite-particle system. Till now, however, there were no (to the best of our knowledge) phenomenological models of a ferrite particle with high-frequency magnetization dynamics that use the effective-mass approximation and the Schrödinger-like equation to analyze *energy eigenstates of a whole ferrite-particle system*, similarly to semiconductor quantum dots.

The recently published theory [2] and new calculation results [3] show that the confined effects for MS oscillations in normally magnetized thin-film ferrite disks are characterized by *discrete energy levels*. As these point (with respect to the external electromagnetic fields) disk particles demonstrate such properties, they should be referred as magnetic dots or magnetic artificial atoms, having the *mesoscopic quantum bound states*. Because of discrete energy eigenstates one can describe the oscillating system as a collective motion of quasiparticles, the light magnons. From the point of view of fundamental studies and new applications, a macroscopic quantum analysis for MS oscillations is very important. In particular it underlies the physics of *magnetoelectric* (ME) oscillating spectrums observed in ferrite disks with surface electrodes [4].

It was supposed in [2] and [3] that the internal DC magnetic field is homogeneous. In this case one can really formulate the spectral problem. At the same time, because of the demagnetizing effects, the internal DC magnetic field in a ferrite disk is essentially non-homogeneous. This should strongly affect on the spectral picture. An analysis of the spectral peak positions for MS oscillations taking into account the DC magnetic field non-homogeneity was made in [5]. Based on an analysis in [5] one cannot, however, determine the character of eigenfunctions. Therefore the

"spectral portrait" of MS oscillations in disks with non-homogeneous internal DC magnetic field becomes unclear. So becomes absolutely unclear the physics of interaction of such a particle with the external RF fields.

In this paper we propose analytical models for enough comprehensive characterization of the spectral properties of magnetic-dipolar-mode oscillations in disk ferromagnetic particles taking into account the DC magnetic field non-homogeneity.

## 2. Spectral properties of MS modes in a ferrite disk with homogeneous DC magnetic field

In a ferrite-disk resonator with a small thickness-to-diameter ratio, the monochromatic MS-wave potential function $\psi$ is represented as [2,3]:

$$\psi = \sum_{p,q} A_{pq} \tilde{\xi}_{pq}(z) \tilde{\varphi}_q(r,\alpha), \tag{1}$$

where $A_{pq}$ is a MS mode amplitude, $\tilde{\xi}_{pq}(z)$ and $\tilde{\varphi}_q(r,\alpha)$ are dimensionless functions describing, respectively, "thickness" ($z$ coordinate) and "in-plane", or "flat" (radial $r$ and azimuth $\alpha$ coordinates) MS modes. For a certain-type "thickness mode" (in other words, for a given quantity $p$), every "flat mode" is characterized by its own function $\tilde{\xi}_q(z)$ and are describing by the Bessel functions.

The spectral problem for MS waves in a ferrite disk resonator can be formulated as the *energy eigenvalue problem* defined by the differential equation:

$$\hat{F}_\perp \tilde{\varphi}_q = E_q \tilde{\varphi}_q \tag{2}$$

together with the corresponding (essential) boundary conditions. A two-dimensional ("in-plane") differential operator $\hat{F}_\perp$ is determined as:

$$\hat{F}_\perp = \frac{1}{2} g \mu \mu_0 \nabla_\perp^2, \tag{3}$$

where $g$ is the unit dimensional coefficient and $\mu$ is a diagonal component of the permeability tensor. The energy orthonormality in a ferrite disk is described as:

$$(E_q - E_{q'}) \int_S \tilde{\varphi}_q \tilde{\varphi}_{q'}^* dS = 0. \tag{4}$$

Because of discrete energy eigenstates of MS-wave oscillations resulting from structural confinement in a case of a normally magnetized ferrite disk, one can consider the oscillating system as a collective motion of quasiparticles – the "light magnons" (lm). One has the following expression for the "light-magnon" average energy of "flat" mode $q$:

$$E_q^{(lm)} = \frac{1}{2} g \mu_0 \left( \beta_q \right)^2, \tag{5}$$

where $\beta_q$ is a MS-wave propagation constant in a ferrite of mode $q$. An effective mass of the light magnon for a monochromatic MS-wave mode is determined as:

$$\left( m_{eff}^{(lm)} \right)_q = \frac{\hbar}{2} \frac{\beta_q^2}{\omega}. \tag{6}$$

The energy levels in a magnetic quantum well and effective masses for the light-magnon modes in a ferrite disk with homogeneous DC magnetic field were calculated in [3]. Calculations were made with use of the disk data given in the Yukawa and Abe paper [5]: saturation magnetization $4\pi M_0 = 1792.7\,G$, disk diameter $2R = 3.98\,mm$, film thickness $h = 0.284\,mm$. The working frequency is $9.51\,GHz$.

## 3. MS modes in a ferrite disk with non-homogeneous DC magnetic field

For a normally magnetized thin-film ferromagnetic disk having the small thickness-to-diameter ratio, the demagnetizing field can be considered just as the radius-dependent function. The internal DC magnetic field is determined in [5] as:

$$H_i(r) = H_0 - H_a - 4\pi M_0 I(r), \tag{7}$$

where $H_0$ and $H_a$ are the applied and the anisotropy fields, respectively, and $I(r)$ is the demagnetizing factor. The function $I(r)$ calculated based on the model described in [5] is shown in Fig. 1 by a solid line.

In this case, however, the standard cylindrical-symmetry problem for MS modes cannot be solved. To understand more clearly this fact let us consider the MS-wave solutions for a general case of an axially magnetized ferrite rod with the permeability-tensor components dependent on a radial coordinate: $\vec{\mu} = \vec{\mu}(r)$. In this case we have the following differential equation for the MS-potential function:

$$\nabla \cdot (\vec{\mu}(r) \cdot \nabla \psi) = 0. \tag{8}$$

It is not difficult to show that in a cylindrical coordinate system this equation has a form:

$$\mu\left(\frac{\partial^2 \psi}{\partial r^2} + \frac{1}{r}\frac{\partial \psi}{\partial r} + \frac{1}{r^2}\frac{\partial^2 \psi}{\partial \theta^2}\right) + \frac{\partial \mu}{\partial r}\frac{\partial \psi}{\partial r} + i\frac{1}{r}\frac{\partial \mu_a}{\partial r}\frac{\partial \psi}{\partial \theta} + \frac{\partial^2 \psi}{\partial z^2} = 0, \tag{9}$$

where $\mu$ and $\mu_a$ are diagonal and off-diagonal components of the permeability tensor. One can see that in this equation separation of variables is impossible and therefore an analytical solution cannot be found. Even assuming possible numerical solutions, it becomes clear, however, that (because of the presence of the *azimuth-first-derivative* term: $\frac{\partial \psi}{\partial \theta}$) these solutions will not be described by single-valued functions.

The absorption peaks are interpreted in [5] to be caused by magnetostatic waves propagating radially across the disk with the DC-field dependent wavenumber in a plane of a YIG film. The mode numbers are determined based on the well-known Bohr-Sommerfeld quantization rule. In definition of the spectral peak positions, the Yukawa and Abe model gives good agreement with the experiments. This fact is illustrated in Fig. 2 for MS modes excited by the homogeneous RF magnetic field (here we use the Yukawa and Abe notation for the mode numbers; in the Yukawa and Abe notation there are odd modes). In spite of so good agreement, one cannot, however, rely on the Yukawa and Abe model as physically describing the experimental situation of interaction of small ferrite disks with the external RF fields. When being excited by the homogeneous RF magnetic field, a small ferrite disk should be considered as a magnetic dipole with evident azimuth variations of the "in-plane" MS-potential function. At the same time, in the Yukawa and Abe model the "in-plane" MS-potential-function distribution is supposed to be azimuthally non-dependent.

## 4. The average procedures for the MS-mode spectral problem in a ferrite disk with non-homogeneous DC magnetic field

The spectral properties of MS modes in a ferrite disk described in [2,3] allow analyzing the real physics of the particle interaction with the external RF fields. In this case, however, the spectral peak positions (found in an assumption of homogeneous internal DC magnetic field and without taking into account the anisotropy field) are rather far from the experimental peak positions. This is illustrated in Fig 2 by squares. The internal field was calculated based on Eq. (7), but with $H_a = 0$ and $I(r) = 1$.

To take into account the internal field non-homogeneity preserving, at the same time, the quantized "spectral portrait" of MS oscillations in a ferrite disk one has to develop certain models of averaging. The initial efforts to create such models one should direct to possible radial averaging the function $\tilde{\mu}(r)$. If such averaging for diagonal and off-diagonal components of the permeability tensor becomes relevant, one can use the separation of variables and obtain, as a result, the spectral characteristics for the oscillating modes. Fig. 3 illustrates typical distribution of $\mu(r)$ calculated based on Eq. (7) with use of the Yukawa and Abe data of the disk parameters and for a certain applied field $H_0$. Distribution of $\mu_a(r)$ has a very similar character and practically coincides with the dependence $\mu(r)$, shown in Fig. 3. From Fig. 3 one can see that the "internal region" – the region where $\mu < 0$ and so the oscillating modes exist – is plateau-like almost in all the range $r < r_1$, where $r_1$ is a the break-radius of $\mu$. The region where $\mu$ becomes sharp dependent on radial coordinate $r$ is very closely abutting to $r_1$. This gives a real possibility to introduce a certain procedure for $\mu$ averaging in a region $r < r_1$. We made $\mu$ averaging for a ferrite disk with an effective diameter $2(s_1 - \delta s_1)$, where $s_1 = \dfrac{r_1}{R}$ is a relative break-radius and $\delta s_1$ is the deviation of a relative break-radius. In our calculation we took $\delta s_1 \leq 0.1 s_1$. For $\mu$ averaged in the above region and in supposition that a ferrite disk has an effective diameter $D_{eff} = 2(s_1 - \delta s_1)$ we solved the spectral problem based on the method described in [2, 3]. The calculation results depicted in Fig. 2 by triangles show good agreement with the Yukawa and Abe experimental data. Fig. 3 gives a concrete picture of the $\mu(r)$ function corresponding to the first mode in Fig. 2. The first mode has smallest diameter $D_{eff}$. The higher mode number, the closer diameter $D_{eff}$ is to a real disk diameter $D = 2R$.

The $\mu$ - averaged calculation results lies down almost completely on a curve calculated based on the Bohr-Sommerfeld quantization rule for MS modes used in [5] (see Fig. 2). This gives us the possibility to make further justification of our $\mu$ - averaged method. The Bohr-Sommerfeld integral for MS modes is [5]:

$$n = \frac{2}{\pi} \frac{D}{h} \int_0^{s_1} \frac{1}{\sqrt{-\mu}} \tan^{-1} \frac{1}{\sqrt{-\mu}} \, ds . \qquad (10)$$

One can see that the integrand aims to zero for $|\mu| \to \infty$. So the role of the abutting region $\delta s_1$ becomes negligibly small in definition of the mode number.

The above procedure of radial averaging the function $\tilde{\mu}(r)$ gives good agreement with the experimental results. Nevertheless, the fact that for different modes one has different effective diameters makes impossible to formulate classical spectral problem, where the domain of definition of a differential operator should be the same for different modes. To solve correctly the spectral problem one has to use another averaging procedure – the averaging the function $I(r)$ in the region of the real disk diameter. Fig. 1 shows the quantity $I(r)_{average}$ based on which the peak positions were calculated. The calculation results shown in Fig. 2 by circles give good agreement with experimental peak positions.

## 5. Conclusion

Based on the above models of averaging, the energy spectra in MS-wave ferrite disks taking into account non-homogeneity of the internal DC magnetic field can be effectively calculated. This makes solvable the problem of interaction of such ferrite particles with the external RF fields.

Figure captions:

Fig. 1. Radial distribution of a demagnetization factor and averaging on the real-disk-diameter region.

Fig. 2. Mode number positions with respect to the applied magnetic field for different calculation methods.

Fig. 3. Averaging procedure for permeability-tensor components on the effective-diameter-region

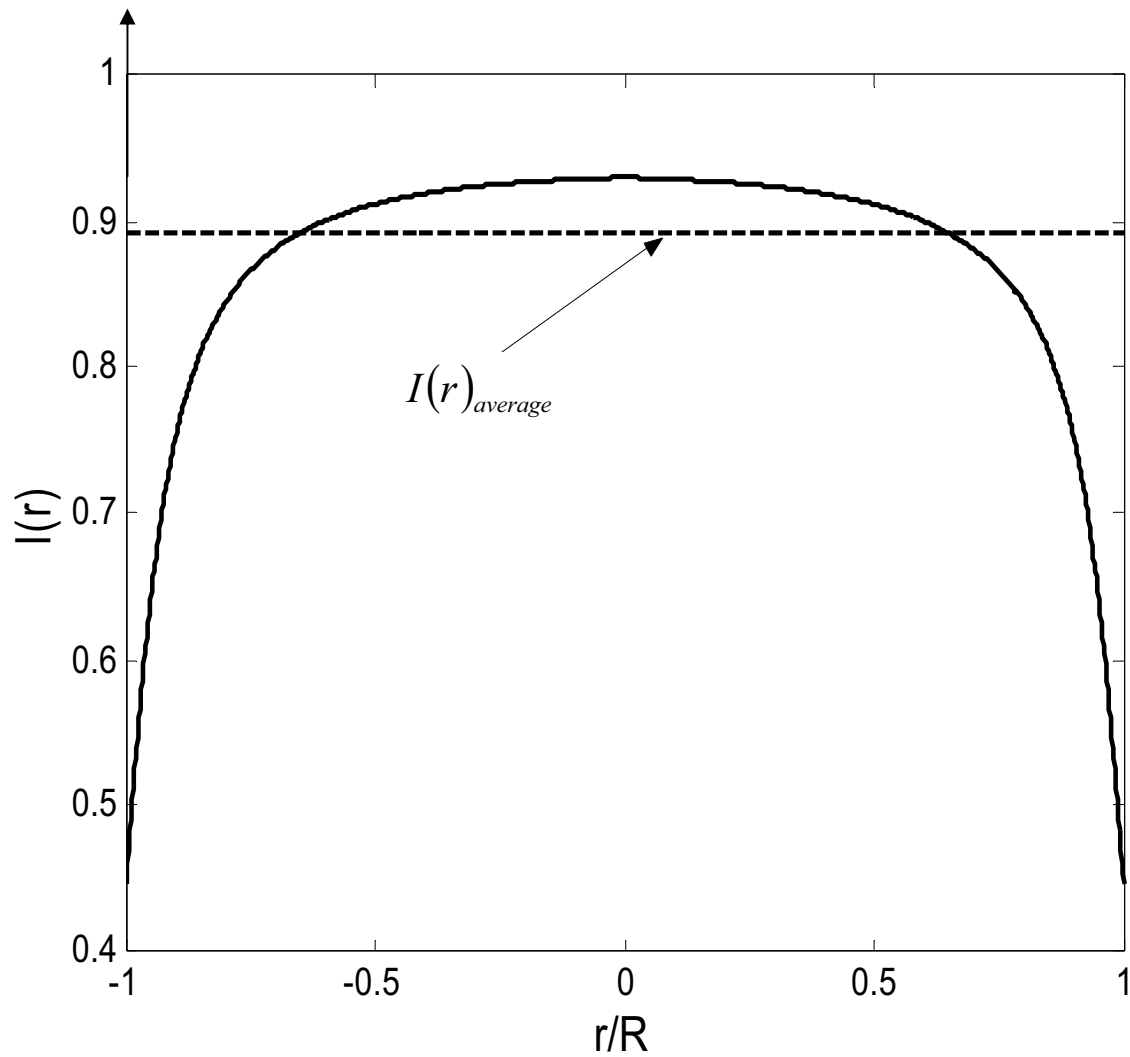

Fig. 1. Radial distribution of a demagnetization factor and averaging on the real-disk-diameter region.

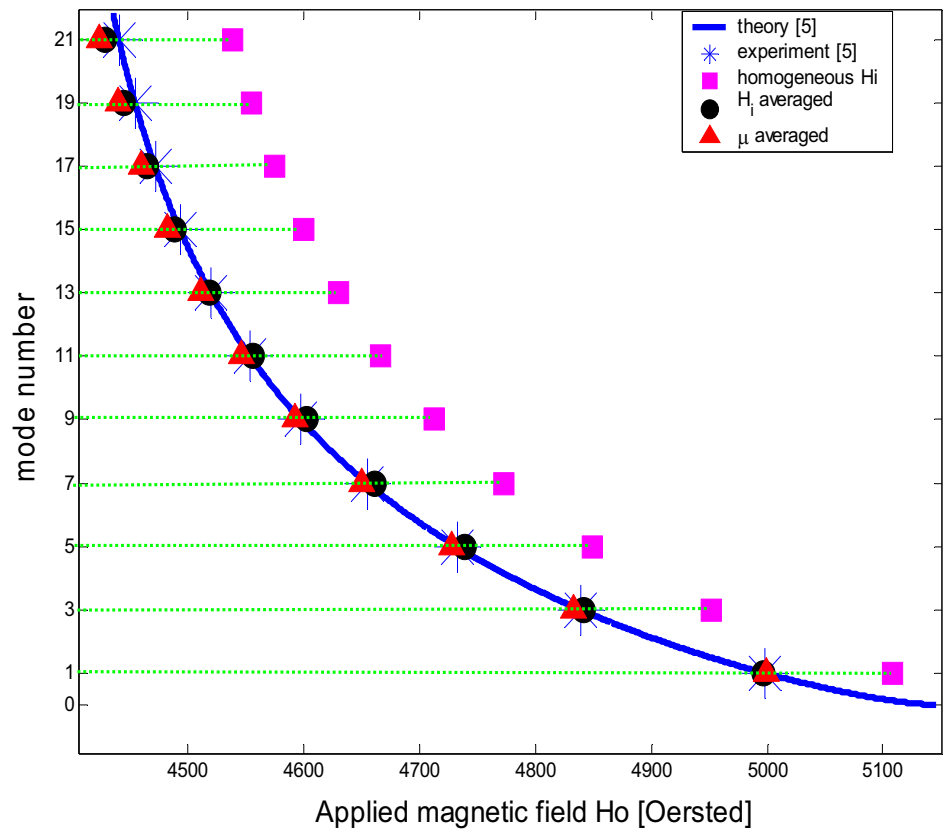

Fig. 2. Mode number positions with respect to the applied magnetic field for different calculation methods.

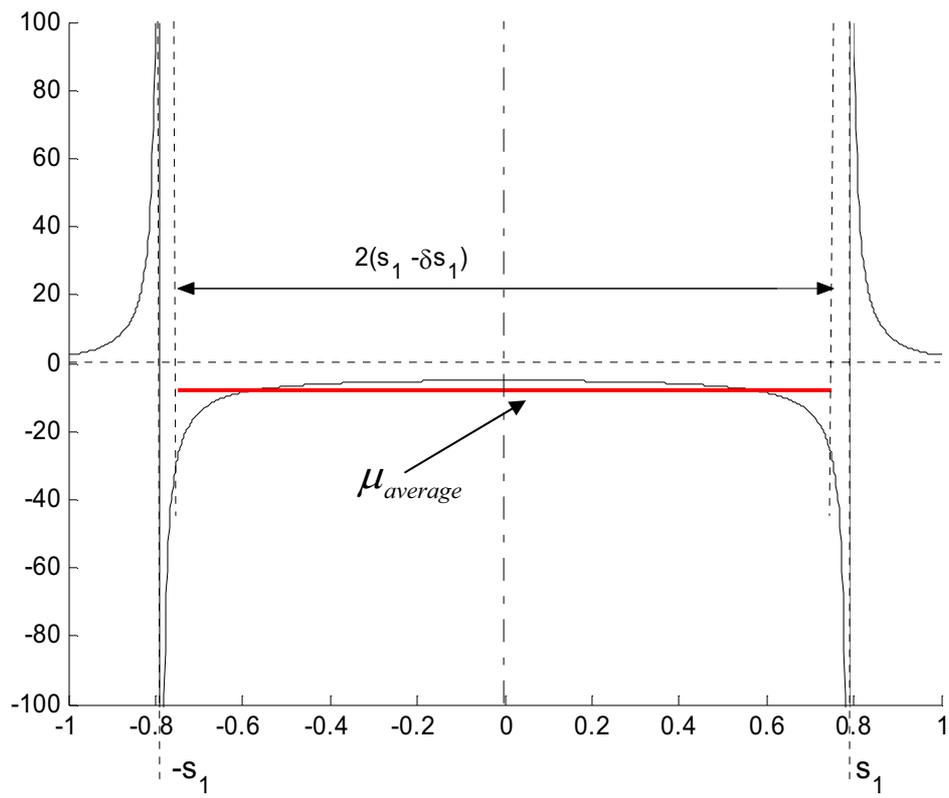

Fig. 3. Averaging procedure for permeability-tensor components on the effective-diameter-region